\documentclass[twocolumn,amsmath,amssymb,aps,prl]{revtex4}

\usepackage{graphicx}
\usepackage{dcolumn}
\usepackage{bm}
\begin{document}
\bibliographystyle{pnas}
\pagestyle{empty}

\title{Modulation of the Hydrophobic Effect through Confinement: A New Model for Chaperonin Action}
\author{Jeremy L. England$^{\dagger}$}
\author{Vijay S. Pande$^{\ddagger}$}
\affiliation{ Department of Physics$^{\dagger}$}
\affiliation{Departments of Chemistry and Structural Biology$^{\ddagger}$\\
Stanford University, Stanford, CA 94305}
\date{\today}

\begin{abstract}  Despite the spontaneity of some \emph{in vitro} protein folding reactions, native folding 
\emph{in vivo} often requires the participation of barrel-shaped multimeric complexes known as chaperonins.  Although it
has long been known that chaperonin substrates fold upon sequestration inside the chaperonin barrel, the precise
mechanism by which confinement within this space facilitates folding remains unknown.  In order to address this
question, we focused on the solvent of the folding reaction and investigated the effect of confinement within a
chaperonin-like cavity on the configurational free energy of water.
 Using a simple theory for the thermodynamics of density and hydrogen bond order fluctuations in liquid water, we
calculated solvent free energies for cavities corresponding to the different conformational states in the ATP-driven
catalytic cycle of the prokaryotic chaperonin GroEL. 
Our findings suggest that one function of chaperonins may be to trap unfolded proteins and subsequently expose them to a
micro-environment in which the hydrophobic effect, a crucial thermodynamic driving force for folding, is enhanced.
\end{abstract}

\maketitle 
\section{Introduction} Efficient production of natively folded protein is an absolute requirement for cell survival.
Although the amino acid sequence of a small polypeptide generally suffices to specify the chain's native structure under
conditions favorable to spontaneous folding \emph{in vitro}, the protein folding problem \emph{in vivo} is more
challenging in several respects. To begin with, the cell must somehow protect the delicate homeostasis of its proteome
against various environmental stresses that wreak havoc with the thermodynamics and kinetics of folding.  Moreover,
folding reactions that take place in the highly crowded intracellular milieu run the risk of never proceeding to
completion because of the tendency of partially unfolded polypeptides to aggregate non-specifically and irreversibly at
high concentrations. And even under more dilute conditions, a large protein that circuitously explores many non-native
kinetic traps along the way to its native state may fold too slowly to be of any use in a biological context.
 
Proteins known as molecular chaperones enable the cell to overcome these obstacles
\cite{Frydman2001chaprev,Hartl2002chaprev}. Chaperones are characterized by their ability to bind the exposed
hydrophobic moieties of non-native polypeptides and guide these wayward chains back onto productive folding pathways.
Perhaps the most intriguing type of chaperone is the chaperonin,  
a barrel-shaped multimeric complex that engulfs and releases its substrates in an ATP-dependent manner
\cite{Horwich1998grorev,Hartl2002chaprev,Horwich2003grorev}.  Chaperonins sub-divide into groups I (eubacterial) and II
(archaebacterial/eukaryotic), and the group I tetradecamer Hsp60/GroEL from \textit{E. coli} is the most extensively
studied model of chaperonin action. The open conformation of GroEL uses its largely hydrophobic inner surface to snare
an unfolded substrate. Subsequent binding of ATP and the GroES co-factor induces a conformational change in GroEL, after
which the substrate may attempt to fold inside a closed complex whose interior cavity surface displays many hydrophilic
residues.  Once ATP hydrolysis takes place, GroES dissociates from the complex and the barrel re-opens, releasing the
substrate into the surroundings and completing the reaction cycle.

For some of its substrates, GroEL mainly provides an infinite-dilution environment in which a protein may fold without
risk of aggregation. For others, however, it has been shown that confinement within the chaperonin barrel is accompanied
by a marked increase in the rate of folding even under conditions that preclude aggregation \cite{Brinker2001twosubstr}. 
The origins of this chaperonin foldase activity are still largely unclear, but recent experiments have begun to shed
some light on the question \cite{Tang2006charge}.  Tang et. al. assayed the activity of a wide range of GroEL mutants
and found 
that the acceleration of folding achieved by a chaperonin mutant correlated with the net charge and hydrophilicity of its
interior cavity surface.  This intriguing result suggests that the hydrophilic walls of the cavity somehow reshape the
folding energy landscape of a confined protein to favor progression to the native state \cite{Horwich2003grorev}.  At
present, the mechanism for this reshaping remains mysterious, and an explanation for how it might come about would
undoubtedly do much to improve our understanding of chaperonin action.

This study aims to provide such an explanation by considering the role of water in folding under confinement. Folding
reactions take place in aqueous solvent, and it is the strong tendency of water molecules to hydrogen bond with each
other that generates the ``hydrophobic effect" \cite{Chandler2005phobrev}, a thermodynamic force that helps drive
proteins to fold by causing non-polar amino acid residues to reduce their solvent-exposed surface area
\cite{Spolar1989phob}. Hydrogen bond quality is
 sensitive to the relative orientations of molecules, and numerous studies have documented substantial changes in the
behavior of water once the liquid is confined in a small enough space that the organization of the hydrogen bond network
becomes significantly constrained \cite{Raviv2001film,Singh2006superphob,Byl2006nanotube}. It is therefore reasonable to
hypothesize that the high degree of confinement experienced by water that participates in folding reactions inside GroEL
may substantially alter the solvent free energy, and therefore the folding landscape, relative to what it would be in
the cytosol. 

Here, we investigate the capacity of a nanopore to modulate the free energy of water confined within it by undergoing
shifts in surface chemistry that mimic the conformational changes in GroEL's catalytic cycle.  Using a phenomenological
theory for the free energy of fluctuations in density and hydrogen bond order in bulk water, we compute the effect of
confinement in hydrophilic and hydrophobic cavities on the ``folding landscape" of a simple model protein. 
Our findings suggest that GroEL may be designed to preferentially bind unfolded substrates and allow them to fold in a
micro-environment in which the hydrophobic effect is enhanced above cytosolic levels.

\section{Model} 
In order to examine the possible role of GroEL in remodeling the solvent of the folding reaction, we sought to develop
an analytical framework for describing the thermodynamics of confined water.
 Microscopic theories of molecular liquids are notoriously difficult, especially for a molecule with water's rare
tendency to form extensive, tetrahedrally-coordinated networks of hydrogen bonds.  For this reason, we chose to avoid
microscopics in favor a phenomenological approach.  Following past work \cite{Tanaka1998LG} that attempted to explain
the anomalous bulk properties of water using a Landau-like statistical field theory, we posited that the effective free
energy of a fluctuation in the configuration of bulk water at a given temperature $T$ and chemical potential $\mu$
depends entirely on two order-parameters: the local density of the fluid $\rho(\mathbf{r})$ and the local quality of the
hydrogen bond network $s(\mathbf{r})$.  Defining $\delta\rho$ and $\delta s$ to be the respective deviations of these
quantities from their average values in the bulk, we can write the free energy of water for small configurational
fluctuations in a volume $V$ as
\begin{multline}
\label{eq:landau}
\mathcal{F}_{V}= \frac{1}{2}\int_{V} d^{3}\mathbf{r} [D_{\rho}(\nabla\rho)^{2}+D_{s}(\nabla s)^{2}+\\
m_{s}\delta s^{2}+m_{\rho}\delta \rho^{2}-\epsilon~\delta s~\delta\rho]+\mathcal{O}(\delta^{3})
\end{multline} The coefficients multiplying each term in the integral should, in principle, be functions of $\mu$ and
$T$, although we would expect the coupling $\epsilon$ between the density and hydrogen bond order to roughly equal the
energy of a hydrogen bond.  The derivative terms simply ensure that both order parameters not vary too rapidly over
space, since they are only well-defined quantities on length scales larger than a water molecule.  The squared terms 
ensure that larger deviations from bulk values of the order parameter carry higher free energetic costs, while the
cross-term proportional to $\delta\rho~\delta s$ dictates that the formation of more hydrogen bonds lowers the energy of
the system.
 
It is important to note here that because we did not include cubic and quartic terms in our free energy functional, we
would be unable to use it to capture the trade-off between fluid density and hydrogen bond quality that leads to the
anomalous decrease in the density of water as it cools to its freezing point. While such effects may, in fact, be
important to adequately describing confinement effects in water, our initial analysis assumes for the sake of simplicity
that the deviations from bulk behavior involved in the phenomenon in question are small enough that the quadratic
functional written above will suffice.

Thus far, we have presented a simple theory of order fluctuations in bulk water. Before we can apply this theory to the
study of confined water, however, we must augment it by introducing an additional contribution to the free energy
functional that comes from the interaction between the liquid and the surface that encloses it.  Thus, let our system
consist of water confined within a shell $S$ of thickness $w$, where $w$ is comparable to the size of a single water
molecule. If $S$ is perfectly non-polar, then it cannot participate in hydrogen bonds and we must require that the
contribution to the free energy due to hydrogen bonding that would be given by  $\epsilon ~\rho~ s$ in the bulk must be
zero in the shell.  We may further introduce some potential $\epsilon ~u(\delta\rho,\delta s)$ that accounts for any
additional effects from the surface chemistry of $S$.  Thus, we obtain a new term in our functional:
\begin{equation}
\mathcal{F}_{S} = w\epsilon\int_{S} dS
\left[\lambda~\rho~ s + u(\rho,s)\right]
\end{equation}
Here, $0\leq\lambda\leq1$ is a parameter that reflects the degree of hydrophobicity of the surface, with $\lambda = 1$
corresponding to a completely non-polar surface.

To find the equilibrium density and hydrogen bond order, we minimize the total free energy
$\mathcal{F}_{T}=\mathcal{F}_{V}+\mathcal{F}_{S}$ with respect to the order parameter fields by using the calculus of
variations. The
 equilibrium fields $\delta\rho_{eq}$ and $\delta s_{eq}$ will obey a system of Helmholtz equations that may be
de-coupled by a change of basis. In the cases of spherically and planar symmetric systems, it is straightforward from
there to obtain closed-form solutions whose integration constants must be set by plugging back into the functional and
minimizing $\mathcal{F}_{T}$.  For arbitrary cavity geometries and surface chemistries, however, it is necessary to seek
a numerical solution. We therefore restrict our attention here to the analytically tractable cases, which are somewhat
easier to deal with and nevertheless suffice as tools for examing the issues of interest in this study.

\section{Results} We modeled GroEL as a spherical pore of radius $R_{G}$ and the chaperonin substrate protein as a
smaller sphere of radius $R_{p}$.  The chaperonin barrel was assumed to have two different states that reflected the
open and closed conformations identified in structural studies of GroEL \cite{Braig2002groxtal}. 
In the open state, the inner surface of the chaperonin was treated as a largely non-polar surface, meaning
$\lambda_{open}=1$ and $u_{open}(\delta\rho,\delta s) = 0$. In the closed state, the chaperonin cavity was treated as
being highly hydrophilic because of its richness in charged 
and polar residues that participate in hydrogen bonds with water molecules. Thus, we set $\lambda_{closed} = 0$ and
$u_{closed} = -\bar{u}~\delta\rho,$ where $\bar{u}>0$ accounts for the net attraction of favorably-oriented molecular
electric dipole moments to the net charge of the cavity surface. The protein was also assumed to have two states: native
and unfolded.  The native conformation was assumed to have a hydrophilic surface chemistry described by
$\lambda_{folded} = u_{folded} = 0$, and the unfolded state was assumed to be partially hydrophobic, with $u_{unfolded}
= 0$ and $\lambda_{unfolded} >0$.  

The solutions to the resulting Helmholtz equations were
 linear combinations of terms of the form $a\exp[\pm c  r]/r$, with the coefficients $a$ determined through minimization
of the total free energy $\mathcal{F}_{T}$, and the correlation length scales $1/c$ set by the parameters of the theory.
In order to plot these solutions, it was necessary to choose specific values for the parameters in the free energy
functional.  Since the questions we sought to answer were qualitative in nature, the precise parameter set used for the
study was somewhat arbitrary.  Indeed, we were unable to observe any qualitative change in the results from varying the
coefficients so long as it was assured that bulk density and hydrogen bond order remained a free energy
\emph{minimum}.  The constraints we ultimately made sure to satisfy were that the shell width $w$ and the characterstic
correlation length scales for the order parameters $1/c$ were comparable to the size of a few water molecules and that
the protein and chaperonin cavity were sized correspondingly.  

\begin{figure}[t]
\resizebox{\columnwidth}{!}{
\includegraphics{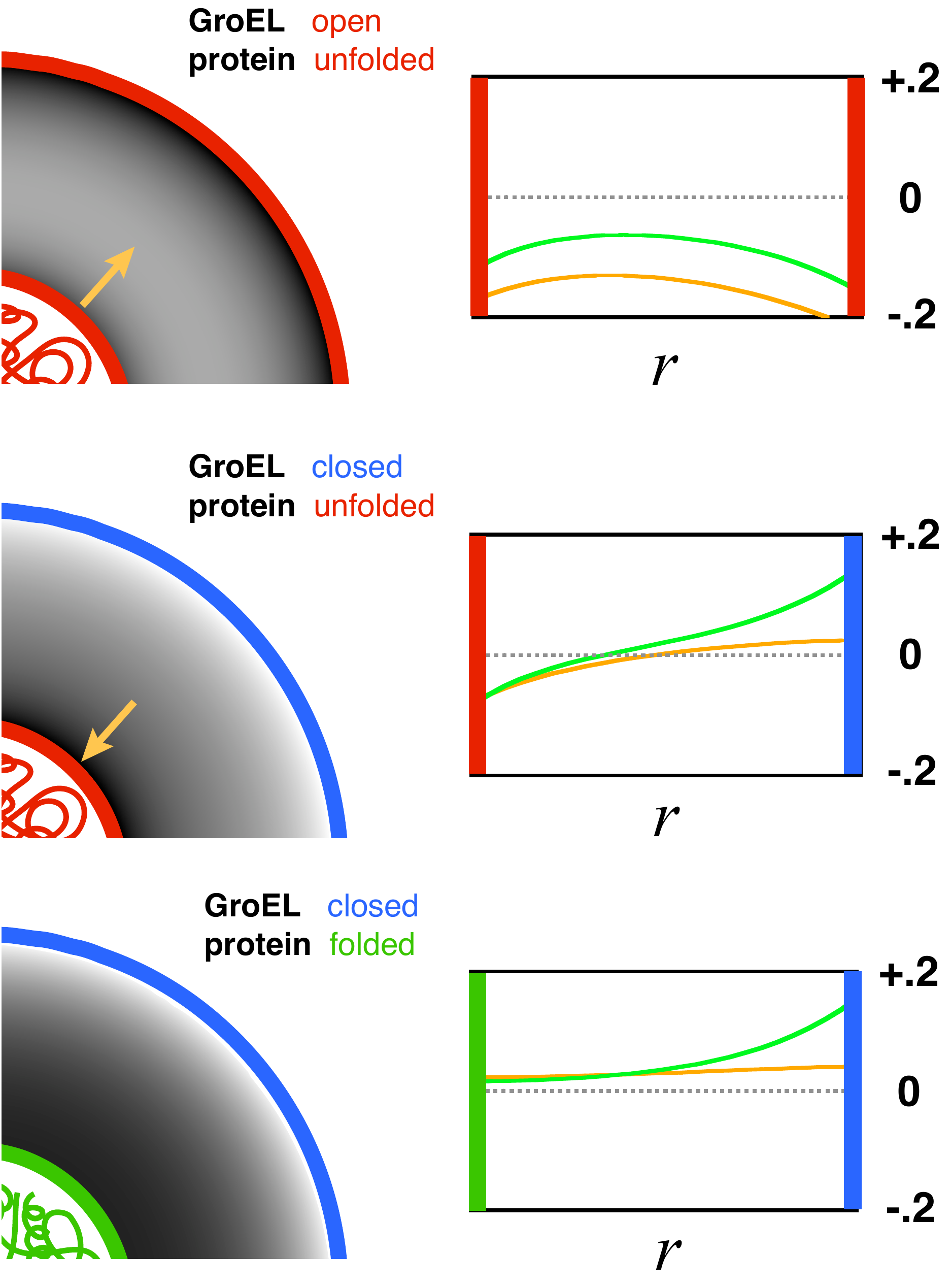} }
\caption{\label{fig:pmf} Equilibrium order parameters $\delta\rho_{eq}$ and $\delta s_{eq}$ inside spherical shells were
calculated using $m_{\rho}=20$, $m_{s}=10$, $D_{\rho}=20$, $D_{s}=30$, $\epsilon = 5$, $\rho_{0}=s_{0}=1$, $w=.5$,
$\bar{u}=1$, $R_{p} = 3$, and $R_{G}=7$.
 All data reported in this study use the same parameters for the free energy functional. 
\emph{Left column}: Liquid density (darker shading corresponding to lesser density) is plotted in the space between the
surface of the protein (red unfolded, green folded) and the cavity (red hydrophobic, blue hydrophilic) wall for an
unfolded protein inside an ``open" hydrophobic cavity (top), an unfolded protein inside a ``closed" hydrophilic cavity
(middle), and a folded protein in a closed cavity (bottom).  Gold arrows indicate the direction of the solvent-mediated
force between the two surfaces (See Fig. \ref{fig:force}). 
\emph{Right column}: Liquid density (green curves) and hydrogen bond order (orange curves) are plotted as a fraction of
their bulk values.  At a hydrophobic surface there is a loss of hydrogen bonding and a depletion of liquid density.  In
contrast, at a highly hydrophilic surface there is an elevation in liquid density and a greater amount of hydrogen
bonding.}
\end{figure}

\begin{figure}[t]
\resizebox{\columnwidth}{!}{
\includegraphics{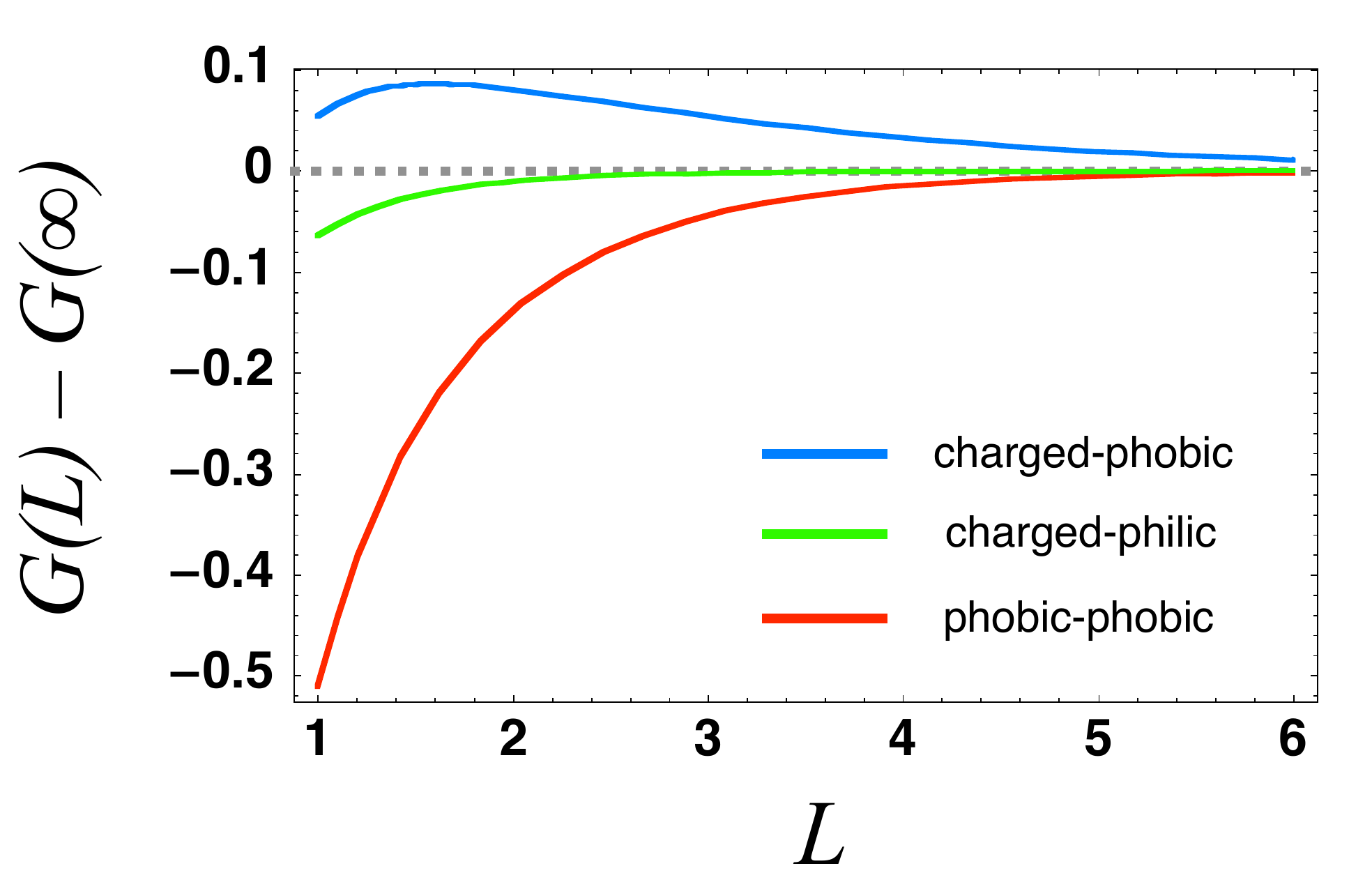} }
\caption{\label{fig:force} Solvent free energy in arbitrary units is plotted as a function of separation $L$ for pairs
of horizontal plates.  Distance is measured in units of the shorter correlation length of the Landau theory, roughly
equivalent to the size of a few water molecules.  Between two hydrophobic plates (red curve) the solvent mediates an
attractive force that grows stronger with proximity.  Between a hydrophobic plate and a highly hydrophilic plate (blue
curve), there is a weaker repulsive force.  Between two hydrophilic plates the force is essentially non-existent (green
curve).}
\end{figure}

Figure \ref{fig:pmf} displays the equilibrium liquid density and hydrogen bond order profiles for the different states
of GroEL and its substrate. For the case of a non-native substrate inside an open GroEL complex, both surfaces confining
the solvent are hydrophobic (Fig. \ref{fig:pmf}, top).  As a result of the loss of hydrogen bonding at these surfaces,
water withdraws from them and from the cavity as a whole, driving the average density below bulk levels.  Moreover, it
is apparent from a calculation of the free energy of solvent trapped between two planar surfaces of the same type that
the force between the non-native substrate and the chaperonin wall is attractive (Fig. \ref{fig:force}, red curve), and
would favor binding.  Once the chaperonin forms a closed complex and 89890undergoes its conformational change,
however, the projection of charged and polar side-chains into the barrel's interior causes the solvent order
and resulting thermodynamic forces to change.
 With the inner surface of the barrel now highly hydrophilic, the solvent density is elevated near the chaperonin wall
and depressed surrounding the unfolded protein (Fig. \ref{fig:pmf}, middle).  The result is a repulsive force between
the surfaces that pushes the substrate towards the center of the cavity (Fig. \ref{fig:force}, blue curve). 
Upon folding, the substrate surface becomes more hydrophilic, allowing the density of the nearby solvent to relax to
bulk levels (Fig. \ref{fig:pmf}, bottom) with the result that the force between chaperonin and substrate becomes very
weak (Fig. \ref{fig:force}, green curve).  

\begin{figure}[t]
\resizebox{\columnwidth}{!}{
\includegraphics{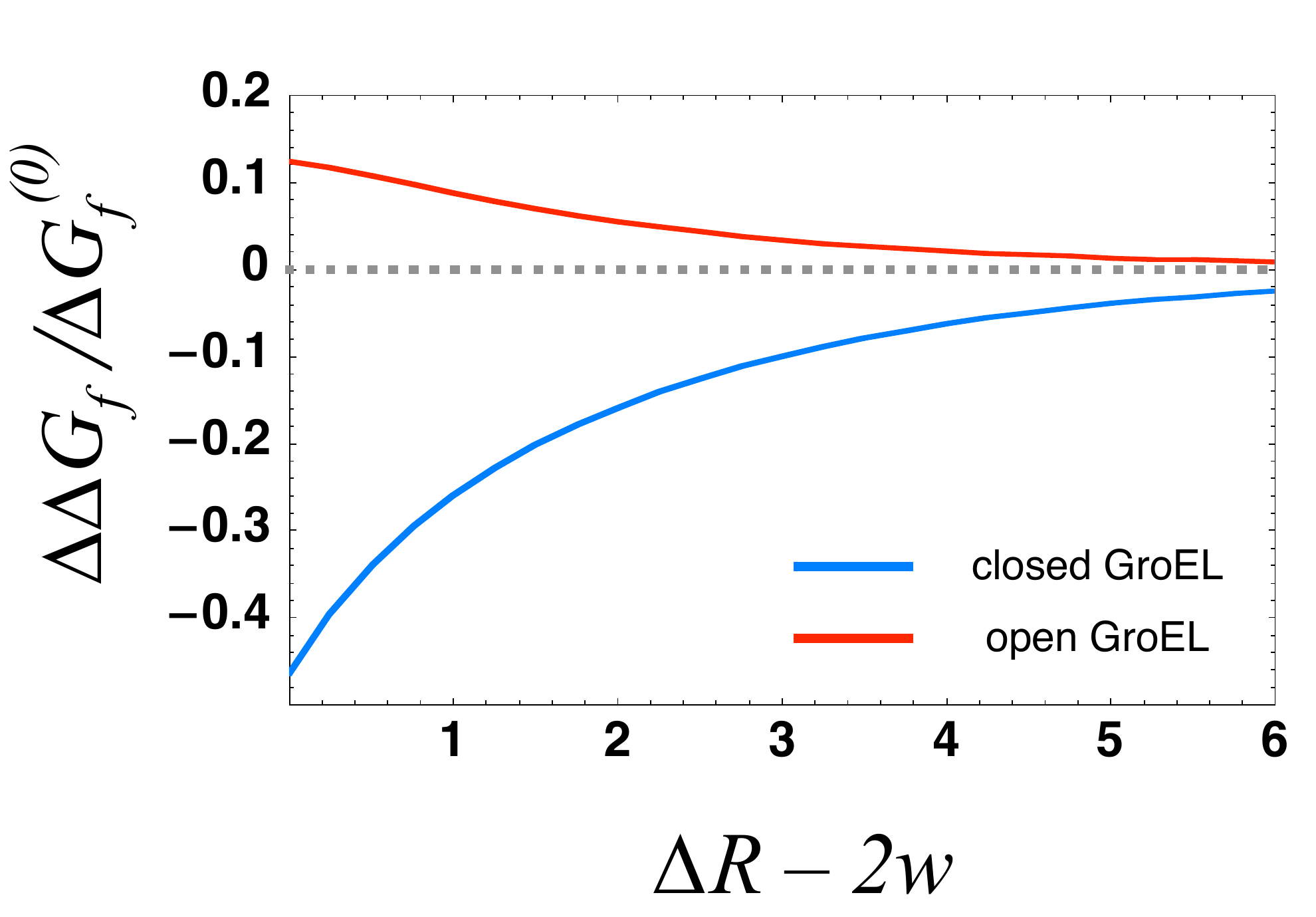} }
\caption{\label{fig:fold} The change in the folding free energy $\Delta\Delta G_{f}$ is defined as the difference
between the free energy of folding under confinement $\Delta G_{f}^{*}$ and the free energy of folding in bulk solution
$\Delta G_{f}^{(0)}$. The fractional change in this solvent free energy of folding is plotted as a function of
confinement radius $\Delta R = R_{G}-R_{p}$ for $R_{p}=3$ less the width $w$ of the interaction shell at each surface. 
Length is measured in units of the shorter correlation length of the theory of solvent fluctuations described in
equation \ref{eq:landau}. When the protein is confined in the largely hydrophobic open GroEL cavity (red curve), the
folded state is destabilized moreso as the degree of confinement increases.  However, when confined in the closed,
highly hydrophilic chaperonin cavity (blue curve), stabilization of the folded state increases with the degree of
confinement.}
\end{figure}

With equilibrium density and bond order profiles in hand, we were able to compute the solvent free energy changes for
folding inside the different conformers of GroEL (Fig. \ref{fig:fold}).  Within the framework of the model
considered here, folding
without the involvement of a chaperonin simply consists of a gradual reduction in hydrophobicity of the surface of a
sphere of radius $R_{p}$ surrounded by aqueous solvent.  In contrast, chaperonin-assisted folding divides into several
stages.  First, the hydrophobic interior of the barrel in its open conformation decreases its free energy by binding to
hydrophobic moeities exposed by a partially unfolded substrate.  Once the non-native protein is inside the cavity, the
solvent free energy
change associated with folding is made less negative by the surrounding hydrophobic walls, and the chaperonin therefore
provides a thermodynamic drive towards unfolding that increases as the radius of confinement becomes smaller (Fig.
\ref{fig:fold}, red curve).
 This effect comes about because the hydrophobic cavity walls deplete
the amount of water in the cavity, making the cost of being unable to form hydrogen bonds at the protein surface lower
than in bulk solution.  After GroEL binds to GroES and ATP and undergoes its conformational change, however, the reverse
is true: the solvent portion of the free energy of folding inside the cavity becomes
\emph{more} negative than that of folding in bulk solution, with the magnitude of the effect again increasing
as the confinement becomes more severe (Fig. \ref{fig:fold}, blue curve). 
This dramatic shift takes place because the charged inner
surface of the chaperonin has elevated the density of the water in the cavity, thereby raising the number of molecules who have difficulty
finding a hydrogen
bonding partner when hydrophobic substrate surface area is presented to the
solvent.
Thus, our model predicts that the conformational change in GroEL brings about a remodeling of the confined solvent that
enhances the hydrophobic effect above its strength in bulk solvent and helps to drive the folding reaction to
completion.

\section{Discussion} Past attempts to explain the foldase activity of chaperonins have pointed to a diverse list of 
possible causes.
 While some researchers have focused on evidence that GroEL may assist in the unfolding of kinetically trapped
substrates \cite{Betancourt1999groel,Cheung2003nanopore}, others see a role for the barrel-shaped cavity in accelerating
folding by reducing the conformational entropy of a protein's unfolded state \cite{Takagi2003confine}, while still
others propose that interactions between the protein and side-chains on the cavity wall might help to smooth out
some of the pitfalls in the protein's free energy landscape \cite{Baumketner2003chap,Jewett2004chap}.  In this work, we
examined a novel dimension of the chaperonin puzzle by studying the thermodynamics of the aqueous solvent in which the
folding reaction takes place.  By developing a simplified theory of order fluctuations in confined water, we were able
compute the free energy of water in cavities that mimicked the salient features of a protein folding inside GroEL.  Our
findings underlined the potential importance of solvent effects in a complete description of how GroEL promotes folding.
\begin{figure}[t]
\resizebox{\columnwidth}{!}{
\includegraphics{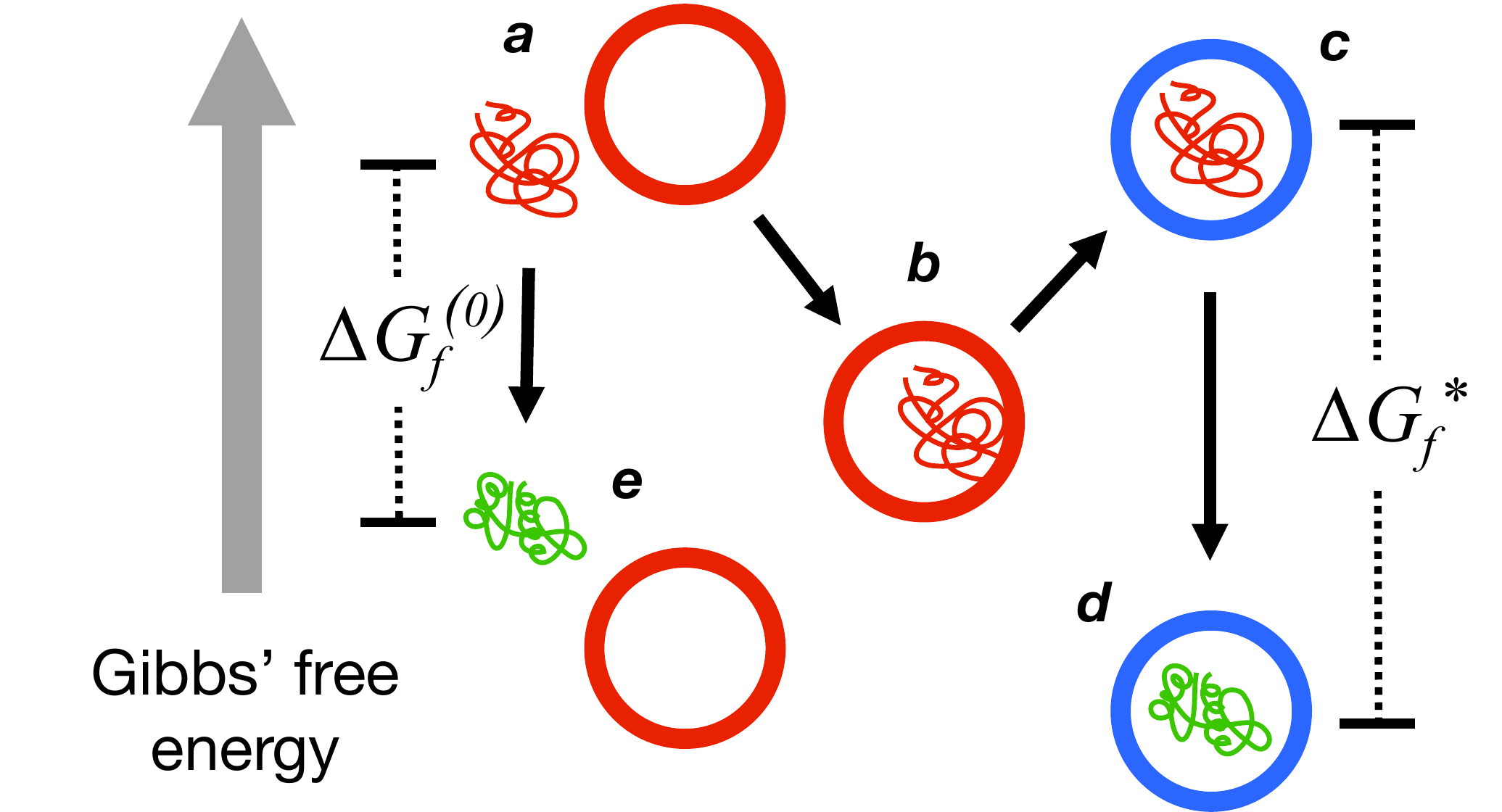} }
\caption{\label{fig:model} A model of chaperonin action.  Hydrophobic forces cause an unfolded (red) protein to bind to
the wall of the open (red) GroEL barrel ($a\rightarrow b$).  Upon formation of a closed (blue) complex containing GroES
and ATP, rearrangements in the barrel present a hydrophilic surface to the interior that repels the substrate into the
center of the cavity ($b\rightarrow c$). Because of this repulsion, the free energy $\Delta G_{f}^{*}$ of folding to the
native 
(green) conformation inside the cavity ($c\rightarrow d$) is more negative than the folding free energy $\Delta
G_{f}^{(0)}$ out in bulk solution ($a\rightarrow e$). The hydrolysis of ATP therefore drives a local enhancement of the
hydrophobic effect inside the chaperonin.}
\end{figure}

Figure 4 summarizes the model of chaperonin action that emerges from our study.  A non-native protein that exposes
hydrophobic groups to the bulk solvent will be engulfed by an open GroEL complex because of hydrophobic forces mediated
by the solvent.  At this point, the substrate becomes bound to the wall of the chaperonin.  Once GroES and ATP bind,
however, the solvent travels uphill in free energy as a conformational change in GroEL presents charged residues to
the barrel's interior.
 As a result, the substrate is forced away from the walls into the center of the cavity, a prediction that is consistent
with experimentally measured increases in substrate fluorescence anisotropy following closing of the complex
\cite{Tang2006charge}.
 The subsequent folding reaction takes place inside a charged, hydrophilic chamber, and is therefore accompanied by a
more drastic decrease in the free energy of the surrounding solvent.  In other words, the interior of GroEL may provide
a micro-environment in which the hydrophobic effect is substantially stronger than in bulk solution.

This finding leads us to propose a novel mechanism for the acceleration of folding by chaperonins.  The hydrophobic
effect is known to be a crucial contributor to the thermodynamic stability of natively folded proteins. However, several
studies have also documented the importance of hydrophobic forces to folding \emph{kinetics}
\cite{Viguera2002phobstab,Northey2002phobstab,Calloni2003phobstab}, both in terms of bringing about nucleation through
hydrophobic collapse \cite{Northey2002phobstab} and through stabilization of the folding transition state
\cite{Viguera2002phobstab}.  It is therefore plausible that the stronger hydrophobic effects experienced by substrates
inside closed GroEL complexes would hasten progression towards the native state.  Thus, through its interaction with the
confined solvent of the folding reaction, the chaperonin could reshape the free energy landscape in order to promote
more rapid folding.

This hydrophobic enhancement model of chaperonin action provides us with a ready means to explain the puzzling
relationship between foldase activity and cavity hydrophilicity reported in recent experiments.  Our model assumes that
charged groups on the interior of GroEL draw water into the barrel chamber, and that this elevation of solvent density
leads directly to an enhancement of hydrophobic effects. The model therefore predicts that replacement of charged amino
acid residues with neutral ones inside the GroEL complex should significantly reduce the chaperonin's foldase activity,
just as was observed in the recent mutational studies of Tang and co-workers \cite{Tang2006charge}. 
Interestingly, the same study also reported a negative correlation between foldase activity and the volume of the GroEL
folding chamber. Although this result was interpreted to be due to the reduction in conformational entropy of the
unfolded protein that confinement causes, it is worth noting that since hydrophobic enhancement rises with the degree of
confinement (Fig. 3), our model provides an alternative explanation for the outcome of those experiments as well.

Additional evidence pointing to charged surface residues as an important determinant of GroEL activity comes from the
work of Wang et. al. \cite{Wang2002optgro}, in which GroEL-GroES was optimized through directed evolution to fold green
fluorescent protein (GFP) more rapidly.  GFP is a small protein with an exceptionally well-buried hydrophobic core that
would be likely to form more rapidly in response to enhancement of hydrophobic effects.  The authors found that a key
mutation for accelerating the folding of GFP was the replacement of an aromatic tyrosine on the interior surface of
GroES with either arginine or histidine, both charged residues.  These results are consistent with our proposal that
charged cavity residues accelerate folding by strengthening the hydrophobic effect.

The existing evidence consistent with the mechanism of chaperonin foldase activity proposed here is substantial, but too
indirect to be compelling by itself. Fortunately, our model's most endearing quality is that it makes clear predictions
that can easily be tested \emph{in silico} and in the laboratory.  At the most basic level, we would expect an all atom
simulation of water and hydrophobic solutes inside a sufficiently hydrophilic cavity to show that hydrophobic forces
between the solutes are enhanced by the confinement.  We would also anticipate that a full-scale simulation of the
closed GroE complex would show that water density inside the cavity was elevated (a prediction already confirmed by
unpublished results), and that the degree of the corresponding elevation in simulations of previously characterized
charged-to-neutral mutants would correlate roughly with their measured foldase activity.
  Finally, and most conclusively, we would predict that experiments that used F\"{o}rster resonance energy transfer
(FRET) or other techniques to measure the strength of hydrophobic forces between solutes inside GroEL would demonstrate
that closure of the barrel through the binding of GroES and ATP is necessary and sufficient to enhance the hydrophobic
effect inside the cavity.

The authors thank E. Miller, D. Kaganovich, D. Lucent for helpful comments  J. England thanks the Fannie and John Hertz
Foundation for support.  This work was also supported by 
NIH NIGMS (R01 GM062868) and the NIH Nanomedicine Center for Protein Folding Machinery (PN1EY016525).

\end{document}


\title{Modulation of the Hydrophobic Effect through Confinement: Supplementary Material}
\author{Jeremy L. England and Vijay S. Pande}
\date{\today}
\maketitle

The free energy functional that must be minimized within the cavity is broken into two components.  The first comes from the bulk
volume term
\begin{equation}
\frac{1}{2}\int d^{3}\mathbf{r}[D_{\rho}(\nabla\rho)^{2}+D_{s}(\nabla s)^{2}+m_{s}\delta s^{2} + m_{\rho}\delta\rho^{2}-\epsilon~\delta s~\delta\rho]
\end{equation}
The second term is the surface interaction term, given by
\begin{equation}
w\epsilon\int_{S}dS[\lambda~\rho~s+u(\rho,s)]
\end{equation}

For symmetrical systems such as planar plates or spherical shells, we proceed by first observing that any fields that are an extremum of the free energy will
satisfy Euler-Lagrange equations of the following form:
\begin{equation}

\end{equation}